\documentclass[11pt]{article}
\usepackage[final]{acl}
\usepackage{times}
\usepackage{latexsym}
\usepackage{graphicx}
\usepackage{booktabs}
\usepackage{amsmath}
\usepackage{multirow}
\usepackage{array}

\title{Context-as-AI-Service: Surfacing Cross-File Dependency Chains\\for LLM-Generated Developer Documentation}

\author{Ameya Gawde, Vyzantinos Repantis, Harshvardhan Singh, Lucy Moys \\
  Meta Platforms, Inc. \\
  \texttt{\{ameyagawde, vrepantis, harshvardhan, ljm\}@meta.com}}

\date{}

\begin{document}
\maketitle

\begin{abstract}
LLM agents increasingly write and maintain developer documentation, but usefulness and accuracy often rely on dependency chains that are not obvious to follow. Even with more files in context, the agent must still decide which cross-file dependencies to trace. We present Context-as-AI-Service (CAIS), a retrieval layer that LLM agents query to find evidence across the codebase as they review or generate documentation. CAIS indexes source code, API references, and upstream documentation, then enables agents to query the index through tool calls that combine keyword and semantic search. We evaluate CAIS in two case studies using Claude Sonnet 4.6 on a production SDK: improving API reference comments in a core source file and validating an LLM-generated tutorial. In both studies, the baseline already had ordinary repository tools such as file reads, keyword search, and symbol navigation. CAIS adds a retrieval layer on top, so the comparison isolates added retrieval rather than basic repository access. In the API-reference review, the CAIS-augmented agent produced the same 5 missing-documentation fixes as the baseline and surfaced 4 findings the baseline missed: 2 cross-file factual errors and 2 underspecified API comments. In the tutorial validation, it surfaced 1 executable bug, 1 API-usage improvement, and 2 missing prerequisites that the baseline pipeline did not catch. These findings required tracing non-obvious dependency chains across utility files, framework internals, usage examples, tests, and component-creation logic. Over five runs per condition, adding CAIS reduced wall-clock time by 22\% to 34\% across the two tasks and lowered input-token usage.
\end{abstract}

\section{Introduction}

Developer documentation is easy to generate but hard to keep correct. API references become stale as code evolves, tutorials accumulate subtle mistakes, and documentation can drift away from the internal codebase that serves as the source of truth for system behavior. LLMs are increasingly used to automate documentation tasks, including generating doc comments, writing tutorials, and maintaining reference pages \cite{chen2021evaluating,nijkamp2023codegen}. LLM-based coding agents can read source files, understand API signatures, and produce fluent prose at scale.

Documentation claims often depend on behavior distributed across files: a method delegating to an internal scheduler, an equality check depending on a separate backing object, a tutorial step relying on framework lifecycle rules, or a component assuming another already exists. Even when many files are available to an agent, the relevant chain may not be obvious. The generated documentation can be locally plausible, yet globally wrong.

Many practical LLM workflows give agents ordinary repository tools such as file reads, keyword search, and nearby-file inspection. This helps when the relevant evidence is local or lexically obvious, but it is less reliable when a documentation claim requires following a non-obvious dependency chain: from an API comment to a registry component, from a method name to a state-management layer, from a tutorial example to a utility constructor, or from an interaction component to framework-level component-creation logic. In these cases, the challenge is not merely having enough tokens. The agent must identify which symbols, examples, tests, or framework rules are worth chasing.

Consider a disposal method documented as removing an object and freeing its resources. From the declaring file, this looks plausible. But the implementation hands the actual removal to a separate state-management layer, where cleanup is deferred until the next processing cycle. The documentation can, therefore, mislead developers about cleanup and synchronization. The defect surfaces only after tracing the dependency from the local API to the layer that defines what removal really does.

We call this the \textbf{cross-file documentation problem}: documentation can look correct in its own file while depending on symbols, implementations, examples, tests, or framework rules elsewhere in the codebase. These errors survive because confirming a claim requires finding the specific files it depends on, and the agent rarely knows in advance which files those are.

To address this problem, we introduce \textbf{Context-as-AI-Service (CAIS)}, a composable retrieval layer for LLM documentation workflows. CAIS does not replace the agent's decision about which files to inspect. It gives the agent a scoped way to retrieve evidence from the codebase before deciding which dependency chains to follow. During documentation generation or review, the agent queries CAIS -- a pre-indexed corpus of source code, API references, tests, examples, and upstream documentation, and receives structured results that point to relevant codebase evidence. The goal is to make non-obvious dependency chains easier to surface without replacing the agent's file inspection or reasoning steps.

This paper makes three contributions:
\begin{enumerate}
    \item We describe a retrieval architecture suitable for documentation workflows, callable by an agent as a tool, that combines keyword and dense search over code, examples, tests, API references, and upstream documentation.
    \item We present two case studies on a production SDK in which this architecture surfaced review findings that baseline workflows with ordinary repository tools missed.
    \item We characterize the cross-file documentation problem and provide an evidence trail showing which dependency chain each retained finding required.
\end{enumerate}

\section{Related Work}

\paragraph{Retrieval-Augmented Generation.}
Retrieval-augmented generation (RAG) grounds LLM outputs in retrieved documents, reducing reliance on parametric memory \cite{lewis2020retrieval}. Prior work has applied retrieval to code tasks such as code generation and summarization \cite{parvez2021retrieval,zhang2023retrieve}. CAIS applies retrieval to a different documentation failure mode: claims that are fluent and locally plausible, but contradicted or incomplete once their codebase dependencies are traced.
\paragraph{Code Documentation Generation.}
Automated documentation generation has progressed from template and heuristic systems \cite{sridhara2010towards,mcburney2014automatic} to neural and LLM-based approaches \cite{hu2018deep,ahmad2020transformer,khan2022automatic,geng2024large}. Much of this work focuses on generating useful or fluent summaries. We focus instead on validation: whether documentation that reads correctly in one file holds up once the relevant cross-file evidence is checked.
\paragraph{LLM Agents for Software Engineering.}
LLM-based software engineering agents can navigate repositories, execute commands, and complete multi-step tasks \cite{yang2024sweagent,jimenez2024swebench}. They typically choose files through repository tools such as file reads, keyword search, and symbol navigation. This works when the relevant evidence is obvious from names, imports, or local call sites, but is less reliable when the dependency chain is semantic or indirect. CAIS complements this agent-driven model with a reusable retrieval layer for surfacing candidate cross-file evidence.

\section{System Architecture}

\begin{figure}[t]
    \centering
    \includegraphics[width=\columnwidth]{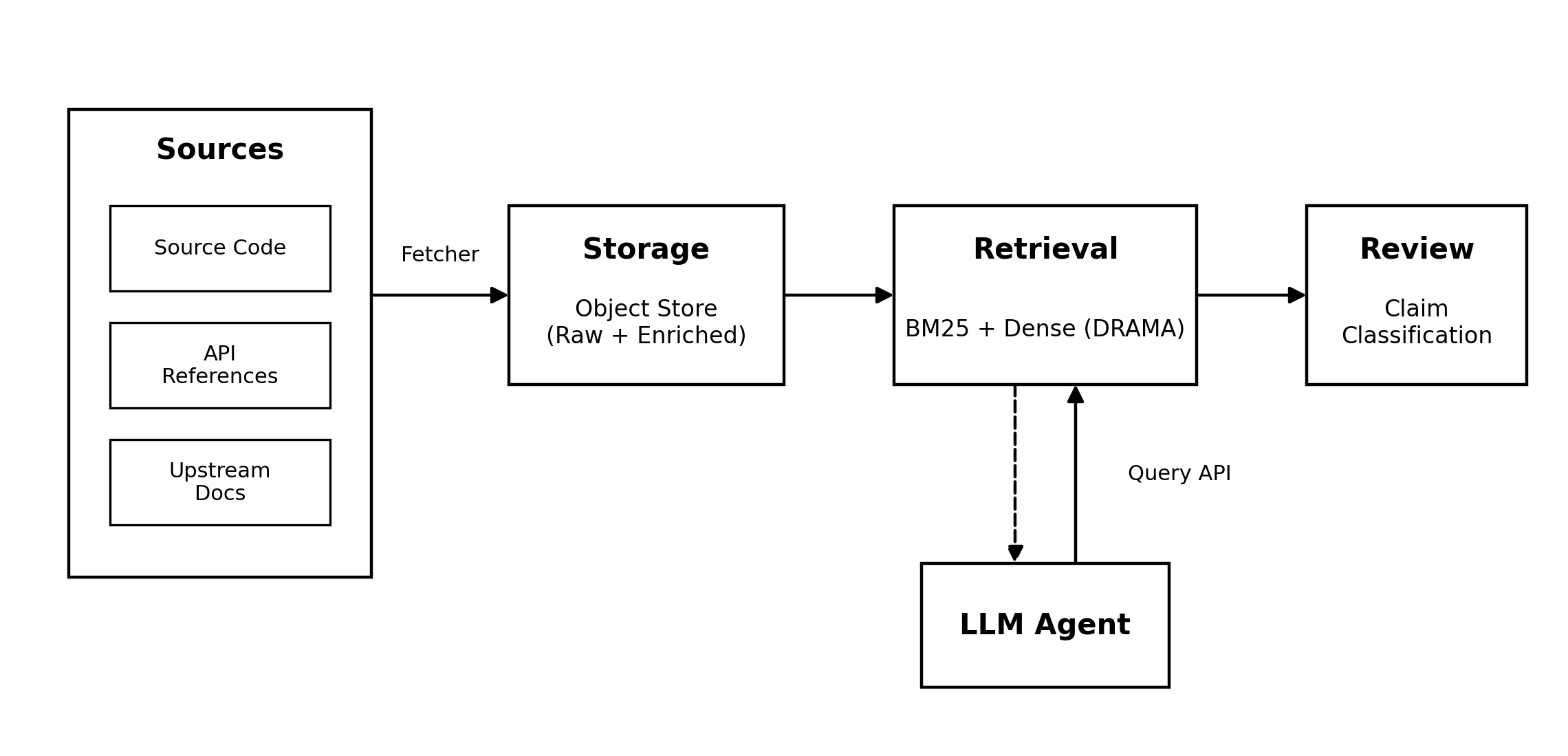}
    \caption{CAIS system architecture. The LLM agent queries a retrieval service during its documentation workflow and receives evidence from across the codebase for documentation review.}
    \label{fig:architecture}
\end{figure}

CAIS is organized as a four-stage pipeline: ingestion, storage, retrieval, and review. Figure~\ref{fig:architecture} illustrates the architecture.

\subsection{Source Ingestion}

The ingestion layer collects three source types:
\begin{itemize}
    \item \textbf{Source code}: SDK implementation files, public APIs, internal utilities, and tests.
    \item \textbf{API references}: Structured documentation extracted from code comments and published API reference pages.
    \item \textbf{Upstream documentation}: Third-party platform or engine documentation relevant to the target SDK.
\end{itemize}

A configurable fetcher normalizes each source into a document record tagged with source type, file path, and last-modified timestamp. This lets one retrieval interface serve code, reference prose, examples, tests, and upstream documentation while preserving enough metadata for freshness checks and re-indexing. The mixed corpus matters because evidence chains often cross source types, such as from an API comment to implementation code, from implementation code to tests, or from tutorial prose to external framework documentation. The corpus itself is configurable, so an operator chooses what to index for a given project. 

\subsection{Storage and Indexing}

Ingested documents are stored in raw and enriched forms. The raw form preserves original formatting for inspection, while the enriched form is tokenized, indexed with BM25, and embedded with DRAMA \cite{ma2025drama}. Because the storage layer supports enumeration and bulk retrieval, the corpus can be re-indexed as the codebase evolves.

CAIS makes project information searchable along both lexical and semantic dimensions, so agents can query for behavior, symbols, examples, temporal dependencies, or failure modes that may not be obvious from the immediate file.

\subsection{Retrieval Interface}

The retrieval component exposes a tool-callable query interface for LLM agents. BM25 handles exact lexical matches such as method names and enum values, while dense retrieval handles semantically related descriptions, examples, and behavioral explanations. At query time, CAIS combines BM25 and DRAMA results using reciprocal rank fusion \cite{cormack2009reciprocal}, then returns ranked snippets with document identifiers and source metadata.

\begin{small}
\begin{verbatim}
Agent -> CAIS: query(
  "deferred removal semantics",
  corpus="codebase", top_k=5)
CAIS -> Agent: [{doc_id: "...",
  score: 0.94, snippet: "object will
  be removed in the next processing
  cycle..."}, ...]
\end{verbatim}
\end{small}

CAIS is agent-agnostic: any LLM agent able to call tools can query it. In our workflows, agents issued partial or behavior-based queries (describing what the code should do, not just its name), retrieved ranked snippets from CAIS, then opened the underlying files when a finding needed broader context. Retrieval guided the agent's exploration. It did not replace reading files, and it did not force the agent to follow a fixed list of queries.

\subsection{Review Layer}

CAIS retrieval results can be used directly by the agent or passed through a lightweight review prompt that labels documentation claims as \textit{consistent}, \textit{contradicted}, or \textit{incomplete}. Retrieved snippets are treated as candidate evidence, not ground truth. In our case studies, a finding was retained only when the evidence supported a concrete correction to the documentation or tutorial.

\section{Case-Study Protocol}

\begin{figure}[t]
    \centering
    \includegraphics[width=\columnwidth]{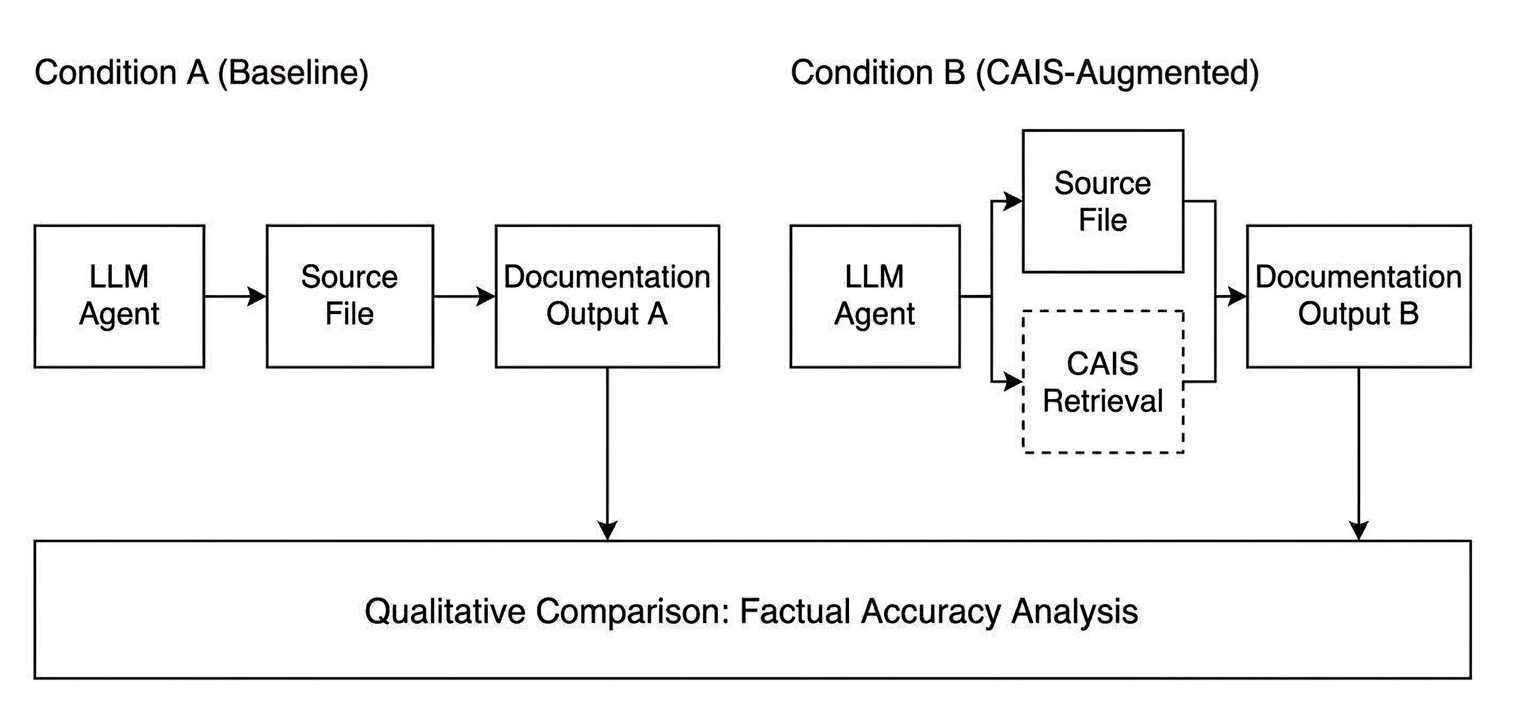}
    \caption{Case-study design. The baseline agent performs the documentation task with ordinary repository tools such as file reads, keyword search, and symbol navigation. The CAIS condition uses the same agent and tools, with retrieval from CAIS added over the indexed SDK corpus. We compare the outputs for extra findings backed by cross-file evidence.}
    \label{fig:evaluation}
\end{figure}

We evaluate CAIS in a retrospective study of two documentation workflows on a production SDK of roughly 200 source files across core, toolkit, and feature packages. Because the SDK and detailed logs are proprietary, we do not try to measure how accurate documentation is across repositories in general. Instead, we ask whether codebase retrieval exposed concrete documentation problems that the recorded baseline workflows missed.

Both studies use the design in Figure~\ref{fig:evaluation}:
\begin{itemize}
    \item \textbf{Baseline}: an LLM coding agent performs the task using its standard repository tools over the working codebase, including file reads, keyword search, and symbol navigation, but without CAIS.
    \item \textbf{+CAIS}: the same agent, prompts, and repository tools, with CAIS retrieval as an extra tool-callable layer over the indexed corpus.
\end{itemize}

The comparison is between using ordinary repository tools alone and using those same tools together with a retrieval layer that helps find cross-file evidence. The question is whether CAIS exposed non-obvious cross-file dependency chains that the baseline workflow did not follow.

We retained a CAIS finding only when three conditions held: (1) the baseline workflow left a meaningful gap -- it missed relations
that matter for correctness or comprehension; (2) the CAIS run drew on context beyond what an LLM would normally see while working on the immediate task; and (3) the finding would concretely affect how an LLM understands the API,
whether the tutorial actually runs, or which implementation steps are needed. Retained findings were manually inspected before inclusion in the evidence trail. This protocol supports a qualitative, evidence-centered case study rather than a controlled benchmark. Section~\ref{sec:evidence} lists each finding, its supporting evidence type, and its developer impact.

\section{Case Studies}

\subsection{Case Study 1: API Reference Improvement}

\paragraph{Task.}
We gave an LLM coding agent a core SDK source file of roughly 900 lines, containing a central domain abstraction, and asked it to review and improve the documentation comments on public members.

\paragraph{Baseline.}
Without CAIS, the agent loaded the source file and its immediate dependencies, identified 5 undocumented public methods, and wrote clean, consistent comments for them. It did not flag any existing comment as wrong.

\paragraph{With CAIS.}
With CAIS access, the agent produced the same missing-documentation fixes and surfaced four additional findings: two cross-file factual errors and two incomplete or underspecified API comments.

First, the documentation for a lookup method referenced a method name renamed in an earlier API revision and called the argument a type identifier. The evidence that CAIS retrieved from a registry component and a query layer showed that the argument is actually an attribute identifier. Second, the documentation for a disposal method implied immediate removal, whereas CAIS retrieved state-management-layer documentation showing that removal is deferred to the next processing cycle. Third, the documentation for an equality method defined equality by object ID alone, but CAIS retrieved implementation and test evidence showing that a separate backing object also matters. Fourth, the documentation for an ownership method largely restated the method name. CAIS retrieved evidence from a filter implementation describing what ownership actually means across multiple clients, the detail needed for a useful comment.

\paragraph{Finding.}
CAIS surfaced non-local evidence needed to evaluate existing comments. The baseline corrected missing comments visible in the file, but missed comments whose correctness depended on registry behavior, state-management semantics, tests, or ownership-filter logic elsewhere in the repository.

\subsection{Case Study 2: Tutorial Generation and Validation}

\paragraph{Task.}
A multi-stage LLM pipeline generated a step-by-step tutorial for an SDK interaction feature. The pipeline researched source code and samples, drafted the tutorial, reviewed the result, validated technical claims against source, and performed a quality critique.

\paragraph{Baseline.}
Without CAIS, the pipeline produced a tutorial covering interaction-subsystem dependencies, creation of an interactive object, the feature's input events, and advanced interaction customization. Its validator confirmed 17 API claims, including enum values, method signatures, component fields, and lifecycle behavior. The tutorial passed all five stages and was marked complete.

\paragraph{With CAIS.}
CAIS-augmented post-generation review surfaced four additional findings. The agent queried CAIS to decide which examples, framework rules, utility APIs, and component-creation behavior to inspect more closely. One finding was an executable bug: the tutorial used an incorrect resource URI scheme for loading external resources. CAIS retrieved canonical examples showing that the scheme prefix already resolves to the resource root, so the tutorial path would double-nest the directory. Because the URI looked syntactically valid, the pipeline's validator (baseline) did not flag it. Another finding was an API-idiom improvement: the tutorial used a verbose scaling constructor even though a shorter equivalent was available in a utility module outside the component's primary definition.

The remaining two findings were critical omissions. First, the tutorial demonstrated two lifecycle callbacks without specifying the required base class, leaving developers without enough information to compile the example. CAIS found this requirement in framework documentation. Second, the tutorial created an interactive object from mesh geometry, but the interaction subsystem auto-generates the needed collision-support component only for primitive shapes. Direct contact interaction could, therefore, fail silently while pointer-based interaction worked, a dependency the baseline's file-by-file validation did not connect to the tutorial scenario.

\paragraph{Finding.}
The baseline validation checked many local API claims correctly, but the retained CAIS findings depended on evidence elsewhere: conventions in example code, a constructor in a utility file, a framework inheritance rule, and a multi-hop component-creation dependency. The failure mode was not a lack of validation, but validation that did not follow the dependency chains needed to tell whether the tutorial would actually work.

\section{Results and Discussion}

\subsection{Summary of Findings}


Table~\ref{tab:results} summarizes the retained findings across both case studies. We separate findings that affect correctness from those that improve documentation quality or use a cleaner API idiom, rather than treating every additional finding as the same kind of error.

\begin{table}[t]
\centering
\small
{\setlength{\tabcolsep}{4pt}%
\begin{tabular}{@{}lcc@{}}
\toprule
\textbf{Finding category} & \textbf{Base} & \textbf{+CAIS} \\
\midrule
Missing public-member docs & 5 & 5 \\
Cross-file factual errors & 0 & 2 \\
Incomplete/underspecified API comments & 0 & 2 \\
Executable tutorial bug & 0 & 1 \\
API-idiom improvement & 0 & 1 \\
Critical omitted prerequisites & 0 & 2 \\
\midrule
\textbf{Total retained findings} & \textbf{5} & \textbf{13} \\
\bottomrule
\end{tabular}}
\caption{Findings retained from the two case studies. CAIS surfaced 8 additional findings not detected in the baseline workflows.}
\label{tab:results}
\end{table}

The baseline findings were missing information available in the local source file. The 8 additional CAIS findings required evidence from registry behavior, scheduling semantics, tests, ownership logic, examples, utility APIs, base-class requirements, and component-creation behavior. Because the baseline already had file reads, lexical search, and symbol navigation, the difference comes from surfacing semantic, cross-file evidence rather than from basic repository access.

\subsection{Evidence Trail}
\label{sec:evidence}

Table~\ref{tab:evidence} provides a compact audit trail for the 8 additional CAIS findings. The table is intentionally phrased in terms of evidence type rather than internal file names, preserving anonymity while showing why each finding required repository-scoped retrieval.

\begin{table*}[t]
\centering
\scriptsize
{\setlength{\tabcolsep}{4pt}%
\begin{tabular}{@{}p{0.04\linewidth}p{0.20\linewidth}p{0.28\linewidth}p{0.23\linewidth}p{0.13\linewidth}@{}}
\toprule
\textbf{ID} & \textbf{Finding} & \textbf{Supporting evidence} & \textbf{Why baseline missed it} & \textbf{Impact} \\
\midrule
F1 & Stale method name and wrong parameter meaning in a lookup method & Registry and query-layer usage showed the argument is an attribute identifier, not a type identifier & Local declaration made the comment look plausible & Wrong API understanding \\
F2 & Disposal method described as immediate removal & State-management layer documented deferred removal in the next processing cycle & Method name suggested immediate cleanup & Cleanup semantics \\
F3 & Equality method described as ID-only equality & Implementation and tests showed equality also depends on a separate backing object & ID-only description looked locally plausible & Multi-client correctness \\
F4 & Ownership method restated the method name & Ownership-filter logic described client-ownership semantics & Comment was vague rather than contradicted & Useful API reference \\
F5 & Incorrect resource URI scheme in tutorial & Canonical examples showed the scheme prefix already resolves to the resource root & URI looked syntactically valid & Tutorial code will fail \\
F6 & Verbose scaling constructor & Utility module provided a shorter equivalent & Constructor lived outside the primary component definition & More idiomatic code \\
F7 & Missing required base class for lifecycle callbacks & Framework documentation specified the class developers must extend & Callback names validated, but inheritance requirement was omitted & Example may not compile \\
F8 & Mesh-based object lacked required collision support & Component-creation logic auto-generates support for primitive shapes but not mesh geometry & File-by-file validation missed the component dependency & Direct contact may fail silently\\
\bottomrule
\end{tabular}}
\caption{Evidence trail for the 8 additional CAIS findings. Each finding required evidence outside the immediate source file or tutorial validation context.}
\label{tab:evidence}
\end{table*}

\subsection{Dependency-Chain Taxonomy}
The case studies suggest a taxonomy by the scope of dependency chain needed to validate a documentation claim:
\begin{enumerate}
    \item \textbf{Single-file}: missing documentation, style issues, or comments the same file contradicts.
    \item \textbf{Cross-file}: stale references, wrong semantics, missing prerequisites, and implicit dependencies that need evidence from other files.
    \item \textbf{Cross-system}: mismatches with upstream platform behavior or version-specific external documentation.
\end{enumerate}
CAIS can retrieve evidence for all three categories, but its added value concentrates in cross-file issues, where the evidence is often not lexically adjacent to the claim, and in cross-system issues when upstream documentation is in the indexed corpus. Single-file issues are already handled well by ordinary LLM workflows, so retrieval adds little there.

\subsection{Efficiency}

To measure efficiency, we ran each condition five times: the agent with its native repository tools, and the same agent with CAIS added. We report mean and standard deviation in Tables~\ref{tab:efficiency-cs1} and~\ref{tab:efficiency-cs2}. The CAIS condition used fewer input tokens and finished faster on both tasks, because retrieval returned compact, pre-ranked snippets for the agent to inspect. LLM calls went up because the agent issued retrieval queries as tool calls. These were queries the agent chose as it went, not a fixed script.

\begin{table}[t]
\centering
\small
\begin{tabular}{@{}lcc@{}}
\toprule
\textbf{Metric} & \textbf{Baseline} & \textbf{+CAIS} \\
\midrule
Wall-clock time (min) & 4.1 $\pm$ 0.7 & 3.2 $\pm$ 0.4 \\
Input tokens (K) & 17.4 $\pm$ 1.8 & 14.6 $\pm$ 1.3 \\
Output tokens (K) & 3.8 $\pm$ 0.6 & 5.1 $\pm$ 0.5 \\
LLM calls & 2.4 $\pm$ 0.5 & 17.6 $\pm$ 0.5 \\
Prompt cache hit & 84 $\pm$ 3\% & 91 $\pm$ 2\% \\
Query latency & --- & 1.6 $\pm$ 0.3\,s \\
CAIS queries & --- & 15.0 $\pm$ 0.0 \\
\bottomrule
\end{tabular}
\caption{Efficiency for Case Study~1, averaged over five runs per condition ($\mu \pm \sigma$). CAIS reduced wall-clock time by 22\% and used fewer input tokens.}
\label{tab:efficiency-cs1}
\end{table}

\begin{table}[t]
\centering
\small
\begin{tabular}{@{}lcc@{}}
\toprule
\textbf{Metric} & \textbf{Baseline} & \textbf{+CAIS} \\
\midrule
Wall-clock time (min) & 17.2 $\pm$ 2.1 & 11.4 $\pm$ 1.3 \\
Input tokens (K) & 112.3 $\pm$ 8.6 & 76.8 $\pm$ 6.2 \\
Output tokens (K) & 9.8 $\pm$ 1.4 & 10.4 $\pm$ 1.2 \\
LLM calls & 17.4 $\pm$ 1.9 & 30.2 $\pm$ 2.4 \\
Prompt cache hit & 71 $\pm$ 4\% & 84 $\pm$ 3\% \\
Query latency & --- & 1.7 $\pm$ 0.3\,s \\
CAIS queries & --- & 13.4 $\pm$ 1.1 \\
\bottomrule
\end{tabular}
\caption{Efficiency for Case Study~2, averaged over five runs per condition ($\mu \pm \sigma$). CAIS reduced wall-clock time by 34\% and used fewer input tokens.}
\label{tab:efficiency-cs2}
\end{table}

\section{Conclusion and Future Work}

These case studies suggest that cross-file documentation inconsistency is a practical failure mode for LLM-assisted documentation workflows. The problem is not only that agents lack enough context; it is that documentation correctness can depend on non-obvious dependency chains that are easy not to trace. CAIS did not merely help generate more text. It exposed contradictions and missing prerequisites whose supporting evidence lived in utility files, examples, tests, framework documentation, and component-creation logic outside the immediate task context.

While our evidence comes from two workflows in one SDK, the observed findings share a common structure: the documentation claim was plausible from the local context, but wrong or incomplete once connected to evidence elsewhere in the codebase. This makes retrieval across the codebase a practical complement to LLM documentation workflows: retrieval can help agents discover which cross-file evidence is relevant before the documentation is trusted.

Several directions remain open. Symbol-level indexing could reduce wasted context-window budget by retrieving specific classes, methods, or tests rather than file-level snippets. Diff-level provenance could help agents reason about when documentation and implementation last changed. Future work should also compare retrieval-augmented review against large-context prompting strategies that explicitly instruct agents to trace dependencies. More broadly, a service like CAIS could serve as a general-purpose context layer for code review, bug detection, and other agent-driven software engineering tasks.

\section*{Limitations}

This work is a case study, not a broad benchmark. We evaluate two documentation workflows on one production SDK, and the findings may not generalize to other languages, repositories, documentation genres, or agent designs. The SDK and full logs are proprietary, so we anonymize identifiers and cannot release the underlying corpus. The evidence trail is therefore self-contained at the level of finding type and retrieved evidence, but not independently reproducible from public data.

Our baseline was a single coding agent (Claude Code v2.1.154 and model Sonnet 4.6) with its native repository tools on one SDK, not an exhaustive set of baselines. A stronger future evaluation would broaden beyond this to alternative retrieval models, different exploration budgets, and large-context models explicitly prompted to trace cross-file dependencies. We also evaluate documentation quality through retained review findings rather than a controlled human-subject study of developer impact. Finally, the current retrieval layer operates mostly at document or file granularity; symbol-level retrieval improves precision and reduces the number of queries needed to reach actionable evidence.

\section*{Ethics Statement}

Our work focuses on improving the accuracy of developer-facing documentation through retrieval-augmented LLM workflows. The case studies used internal, non-public SDK source code with authorization from the codebase owners. No human participants were recruited, no crowd-sourced annotations were collected, and no personally identifiable information was accessed or generated as part of the study.

The LLM agent used in the case studies was accessed through a licensed API under the provider's standard terms of service. We anonymize internal system names, file paths, and identifiable product details, and we present CAIS as a review aid rather than a replacement for human approval.

Automated documentation tools can affect technical writing and developer education workflows. In the deployment context studied here, the bottleneck was documentation freshness and correctness rather than writing capacity; CAIS was used to surface evidence for review, not to remove human oversight.

\bibliography{references}

@article{chen2021evaluating,
  title={Evaluating Large Language Models Trained on Code},
  author={Chen, Mark and Tworek, Jerry and Jun, Heewoo and Yuan, Qiming and
    Pinto, Henrique Ponde de Oliveira and Kaplan, Jared and Edwards, Harri and
    Burda, Yuri and Joseph, Nicholas and Brockman, Greg and Ray, Alex and
    Puri, Raul and Krueger, Gretchen and Petrov, Michael and Khlaaf, Heidy and
    Sastry, Girish and Mishkin, Pamela and Chan, Brooke and Gray, Scott and
    Ryder, Nick and Pavlov, Mikhail and Power, Alethea and Kaiser, Lukasz and
    Bavarian, Mohammad and Winter, Clemens and Tillet, Philippe and
    Such, Felipe Petroski and Cummings, Dave and Plappert, Matthias and
    Chantzis, Fotios and Barnes, Elizabeth and Herbert-Voss, Ariel and
    Guss, William Hebgen and Nichol, Alex and Paino, Alex and Tezak, Nikolas and
    Tang, Jie and Babuschkin, Igor and Balaji, Suchir and Jain, Shantanu and
    Saunders, William and Hesse, Christopher and Carr, Andrew N. and
    Leike, Jan and Achiam, Josh and Misra, Vedant and Morikawa, Evan and
    Radford, Alec and Knight, Matthew and Brundage, Miles and Murati, Mira and
    Mayer, Katie and Welinder, Peter and McGrew, Bob and Amodei, Dario and
    McCandlish, Sam and Sutskever, Ilya and Zaremba, Wojciech},
  year={2021},
  archivePrefix={arXiv},
  eprint={2107.03374},
  journal={arXiv preprint arXiv:2107.03374},
  primaryClass={cs.LG},
  doi={10.48550/arXiv.2107.03374},
  url={https://arxiv.org/abs/2107.03374}
}

@inproceedings{nijkamp2023codegen,
  title={{CodeGen}: An Open Large Language Model for Code with Multi-Turn Program Synthesis},
  author={Nijkamp, Erik and Pang, Bo and Hayashi, Hiroaki and Tu, Lifu and
    Wang, Huan and Zhou, Yingbo and Savarese, Silvio and Xiong, Caiming},
  booktitle={Proceedings of the Eleventh International Conference on Learning Representations},
  year={2023},
  url={https://openreview.net/pdf?id=iaYcJKpY2B_}
}

@inproceedings{lewis2020retrieval,
  title={Retrieval-Augmented Generation for Knowledge-Intensive {NLP} Tasks},
  author={Lewis, Patrick and Perez, Ethan and Piktus, Aleksandra and Petroni, Fabio and
    Karpukhin, Vladimir and Goyal, Naman and K{\"u}ttler, Heinrich and Lewis, Mike and
    Yih, Wen-tau and Rockt{\"a}schel, Tim and Riedel, Sebastian and Kiela, Douwe},
  booktitle={Advances in Neural Information Processing Systems 33 (NeurIPS 2020)},
  editor={H. Larochelle and M. Ranzato and R. Hadsell and M.F. Balcan and H. Lin},
  pages={9459--9474},
  volume={33},
  publisher={Curran Associates, Inc.},
  year={2020},
  doi={10.5555/3495724.3496517},
  url={https://proceedings.neurips.cc/paper_files/paper/2020/file/6b493230205f780e1bc26945df7481e5-Paper.pdf}
}

@inproceedings{zhang2023retrieve,
  title={Syntax-Aware Retrieval Augmented Code Generation},
  author={Zhang, Xiangyu and Zhou, Yu and Yang, Guang and Chen, Taolue},
  booktitle={Findings of the Association for Computational Linguistics: EMNLP 2023},
  pages={1291--1302},
  year={2023},
  publisher={Association for Computational Linguistics},
  doi={10.18653/v1/2023.findings-emnlp.90},
  url={https://aclanthology.org/2023.findings-emnlp.90/}
}

@inproceedings{parvez2021retrieval,
  title={Retrieval Augmented Code Generation and Summarization},
  author={Parvez, Md Rizwan and Ahmad, Wasi and Chakraborty, Saikat and Ray, Baishakhi and Chang, Kai-Wei},
  booktitle={Findings of the Association for Computational Linguistics: EMNLP 2021},
  pages={2719--2734},
  year={2021},
  publisher={Association for Computational Linguistics},
  address={Punta Cana, Dominican Republic},
  doi={10.18653/v1/2021.findings-emnlp.232},
  url={https://aclanthology.org/2021.findings-emnlp.232/}
}

@inproceedings{sridhara2010towards,
  title={Towards Automatically Generating Summary Comments for {Java} Methods},
  author={Sridhara, Giriprasad and Hill, Emily and Muppaneni, Divya and Pollock, Lori and Vijay-Shanker, K.},
  booktitle={Proceedings of the 25th {IEEE/ACM} International Conference on Automated Software Engineering},
  pages={43--52},
  year={2010},
  publisher={ACM},
  doi={10.1145/1858996.1859006},
  url={https://dl.acm.org/doi/10.1145/1858996.1859006}
}

@inproceedings{mcburney2014automatic,
  title={Automatic Documentation Generation via Source Code Summarization of Method Context},
  author={McBurney, Paul W. and McMillan, Collin},
  booktitle={Proceedings of the 22nd International Conference on Program Comprehension},
  pages={279--290},
  year={2014},
  publisher={ACM},
  address={Hyderabad, India},
  doi={10.1145/2597008.2597149},
  url={https://dl.acm.org/doi/10.1145/2597008.2597149}
}

@inproceedings{hu2018deep,
  title={Deep Code Comment Generation},
  author={Hu, Xing and Li, Ge and Xia, Xin and Lo, David and Jin, Zhi},
  booktitle={Proceedings of the 26th International Conference on Program Comprehension},
  pages={200--210},
  year={2018},
  publisher={ACM},
  doi={10.1145/3196321.3196334},
  url={https://dl.acm.org/doi/10.1145/3196321.3196334}
}

@inproceedings{ahmad2020transformer,
  title={A Transformer-based Approach for Source Code Summarization},
  author={Ahmad, Wasi Uddin and Chakraborty, Saikat and Ray, Baishakhi and Chang, Kai-Wei},
  editor={Jurafsky, Dan and Chai, Joyce and Schluter, Natalie and Tetreault, Joel},
  booktitle={Proceedings of the 58th Annual Meeting of the Association for Computational Linguistics},
  pages={4998--5007},
  month=jul,
  year={2020},
  address={Online},
  publisher={Association for Computational Linguistics},
  doi={10.18653/v1/2020.acl-main.449},
  url={https://arxiv.org/abs/2005.00653}
}

@inproceedings{khan2022automatic,
  title={Automatic Code Documentation Generation Using {GPT-3}},
  author={Khan, Junaed Younus and Uddin, Gias},
  booktitle={Proceedings of the 37th {IEEE/ACM} International Conference on Automated Software Engineering},
  year={2022},
  publisher={ACM},
  address={Rochester, MI, USA},
  doi={10.1145/3551349.3559548},
  url={https://dl.acm.org/doi/10.1145/3551349.3559548}
}

@inproceedings{geng2024large,
  title={Large Language Models are Few-Shot Summarizers: Multi-Intent Comment Generation via In-Context Learning},
  author={Geng, Mingyang and Wang, Shangwen and Dong, Dezun and Wang, Haotian and
    Li, Ge and Jin, Zhi and Mao, Xiaoguang and Liao, Xiangke},
  booktitle={Proceedings of the 46th International Conference on Software Engineering},
  pages={39:1--39:13},
  year={2024},
  publisher={ACM},
  address={Lisbon, Portugal},
  doi={10.1145/3597503.3608134},
  url={https://dl.acm.org/doi/10.1145/3597503.3608134}
}

@inproceedings{yang2024sweagent,
  title={{SWE}-agent: Agent-Computer Interfaces Enable Automated Software Engineering},
  author={Yang, John and Jimenez, Carlos E. and Wettig, Alexander and Lieret, Kilian and
    Yao, Shunyu and Narasimhan, Karthik and Press, Ofir},
  booktitle={Advances in Neural Information Processing Systems 37 (NeurIPS 2024)},
  editor={A. Globerson and L. Mackey and D. Belgrave and A. Fan and U. Paquet and J. Tomczak and C. Zhang},
  pages={50528--50543},
  volume={37},
  publisher={Curran Associates, Inc.},
  year={2024},
  url={https://proceedings.neurips.cc/paper_files/paper/2024/file/5a7c947568c1b1328ccc5230172e1e7c-Paper-Conference.pdf}
}

@inproceedings{jimenez2024swebench,
  title={{SWE}-bench: Can Language Models Resolve Real-World {GitHub} Issues?},
  author={Jimenez, Carlos E. and Yang, John and Wettig, Alexander and Yao, Shunyu and
    Pei, Kexin and Press, Ofir and Narasimhan, Karthik},
  booktitle={Proceedings of the Twelfth International Conference on Learning Representations},
  year={2024},
  url={https://openreview.net/forum?id=VTF8yNQM66}
}

@article{ma2025drama,
  title     = {{DRAMA}: Diverse Augmentation from Large Language Models to Smaller Dense Retrievers},
  author    = {Ma, Xueguang and Lin, Xi Victoria and Oguz, Barlas and Lin, Jimmy and Yih, Wen-tau and Chen, Xilun},
  journal   = {arXiv preprint arXiv:2502.18460},
  year      = {2025},
  url       = {https://arxiv.org/abs/2502.18460}
}

@inproceedings{cormack2009reciprocal,
  author    = {Cormack, Gordon V. and Clarke, Charles L. A. and B{\"u}ttcher, Stefan},
  title     = {Reciprocal rank fusion outperforms Condorcet and individual rank learning methods},
  booktitle = {Proceedings of the 32nd International ACM SIGIR Conference on Research and Development in Information Retrieval},
  pages     = {758--759},
  year      = {2009},
  organization={ACM},
  url       = {https://doi.org/10.1145/1571941.157211}
}

\end{document}